# de Haas-van Alphen effect of correlated Dirac states in kagome metal Fe$_3$Sn$_2$

Linda Ye[1], Mun K. Chan[2], Ross D. McDonald[2], David Graf[3], Mingu Kang[1], Junwei Liu[4], Takehito Suzuki[1], Riccardo Comin[1], Liang Fu[1], Joseph G. Checkelsky[1]

[1]*Department of Physics, Massachusetts Institute of Technology, Cambridge, MA 02139, USA*
[2]*National High Magnetic Field Laboratory, LANL, Los Alamos NM 87545, USA*
[3]*National High Magnetic Field Laboratory, Tallahassee, FL 32310, USA*
[4]*Department of Physics, Hong Kong University of Science and Technology, Clear Water Bay, Hong Kong, China*

**The field of topological electronic materials has seen rapid growth in recent years (*1, 2*), in particular with the increasing number of weakly interacting systems predicted and observed to host topologically non-trivial bands (*3-5*). Given the broad appearance of topology in such systems, it is expected that correlated electronic systems should also be capable of hosting topologically non-trivial states (*6*). Interest in correlated platforms is heightened by the prospect that collective behavior therein may give rise to new types of topological states and phenomena not possible in non-interacting systems (*7*). However, to date only a limited number of correlated topological materials have been definitively reported due to both the complexity in calculation of their electronic properties and the experimental challenge of disentangling underlying correlation effects from topological aspects of their electronic structure. Here, we report a de Haas-van Alphen (dHvA) study of the recently discovered kagome metal $Fe_3Sn_2$ mapping the massive Dirac states strongly coupled to the intrinsic ferromagnetic order. We observe a pair of quasi-two-dimensional Fermi surfaces arising from the massive Dirac states previously detected by spectroscopic probes (*8, 9*) and show that these band areas and effective masses are systematically modulated by the rotation of the ferromagnetic moment. Combined with measurements of Berry curvature induced Hall conductivity, we find that along with the Dirac fermion mass, velocity, and energy are suppressed with rotation of the moment towards the kagome plane. These observations demonstrate that strong coupling of magnetic order to electronic structure similar to that observed in elemental ferromagnets (*10-13*) can be extended to topologically non-trivial electronic systems, suggesting pathways for connecting topological states to robust spintronic technologies (*14*). Additionally, our bulk thermodynamic observations provide crucial information to constrain theoretical modeling of magnetic topological metals (*8, 9, 15-17*) and provide insights to realizing further highly correlated topological materials (*18*).**

The two-dimensional kagome lattice is a system of corner-sharing triangles assembled in a hexagonal fashion analogous to the graphene lattice (*19*). These triangular and hexagonal structural features are a test ground to access the physics of magnetic frustration (*20*) and of lattice-driven Dirac fermions (*21*), respectively. In the context of electronic hopping models, the kagome network also gives rise to a flat band together with a pair of Dirac bands that potentially support exotic phases such as interaction-driven ferromagnetism (*22*) and chiral superconductivity (*23*). In reciprocal space, the two Dirac band touching points on the kagome lattice are positioned at the *K* and *K'* points at the Brillouin zone boundary, identical to the case for the graphene lattice and likewise are protected by crystallographic symmetries (*19*). Compared to the graphene lattice, the nearest neighbor bonds in the kagome lattice are not contained in mirror planes perpendicular to the basal lattice and can therefore experience an electric field orthogonal to the nearest neighbor bonds (*24*), introducing explicitly spin-orbit effects into the band structure and pathways to topologically nontrivial electronic bands (*25*).

In terms of material realizations, a number of recent efforts have focused on metallic kagome lattice materials that potentially connect to the electronic hopping behavior expected for the 2D lattice. In particular, the kagome lattice has been realized in a series of hexagonal 3*d* transition metal stannides and germannides (*26, 27*). The basic building blocks consist of a transition metal kagome layer $T_3$(Ge,Sn) with Sn/Ge at the hexagon center, together with a stanene layer (Ge,Sn)$_2$ (see Fig. 1a), forming compounds with chemical formula $[T_3(Ge,Sn)]_x[(Ge,Sn)_2]_y$. Two representative materials, $Mn_3Sn$ ($x = 1, y = 0$) (*15*) and $Fe_3Sn_2$ ($x = 2, y = 1$) (*8*) have recently been identified as hosts to 3D Weyl fermions and quasi-2D massive Dirac fermions, respectively, suggesting the dimensionality of electronic topology is sensitive to the crystallographic arrangement of the basic building blocks, attributed to covalent and metallic bonding of the stanene and kagome layers, respectively (*28*). Moreover, the use of the 3*d* transition elements allows the introduction of magnetism; in the case of $Fe_3Sn_2$ this is a soft ferromagnetic order (*29*) which along with atomic spin-orbit coupling gives rise to substantial intrinsic anomalous Hall conductivity from the massive Dirac bands extending above room temperature (*8*). Given the softness of this magnetic order, a natural question which arises is how a general positioning of the magnetic moment $\vec{m}$ affects the electronic structure and topology of this system. Here we report a torque magnetometry study that captures the evolution of the quasi-2D Dirac bands via the de Haas-van Alphen effect (dHvA) while at the same time monitoring changes in the magnetic order. These observations together demonstrate a systematic development of the massive Dirac states consistent with a Kane-Mele spin-orbit coupling (*30*) with a relativistic energy shift comparable to those observed in elemental ferromagnets (*31*).

Measurement of the magnetic torque $\tau$ for $Fe_3Sn_2$ up to 65 T are shown in Fig. 1(b) for two different angles $\theta_1 = 15°$ and $60°$ for the applied field *H* relative to the *c*-axis of the crystal (see lower panel Fig. 1(a)). For both angles, an initial rise with $H$ gives way to a gradual decay while a sharp low field peak emerges for $\theta_1 = 15°$. As we return to below, this corresponds to the polarizing process of the soft ferromagnetic moment along the field direction, with $\vec{m}$ being aligned along $H$ with a deviation less than 0.02 $\mu_B$/f.u. above 10 T. For temperature $T = 0.4$ K, as $H$ increases above approximately 20 T we see the onset of dHvA oscillations. As shown for $\theta_1 = 15°$, at higher $T = 15$ K the oscillations are suppressed while the overall shape of $\tau(H)$ remains relatively unchanged. This is indicative of the lower energy scale for Landau quantization compared to the magnetic order and associated anisotropy (see supplementary materials). Fig. 1(c) shows the oscillatory component in the transverse magnetization $\Delta M_T \equiv \Delta\tau/\mu_0 H$ ($\Delta\tau$ is the oscillatory part of torque after subtracting a polynomial background) as a function of inverse applied field $\Delta M_T(H^{-1})$ at $T = 0.5$ K for various $\theta_2$. Multiple frequencies are evident accompanied by an increase in frequency of the slowest oscillation with increasing $\theta_2$ (black arrows in Fig. 1(c) trace its 8th and 9th Landau level).The magnitude of the oscillations are consistent with a bulk origin (surface state oscillations would correspond to an amplitude of approximately 4 $\mu_B$ per surface unit cell, comparable with $|\vec{m}|$ itself (*32*)).

The dHvA spectrum for samples A-D at the base $T = 0.5$-$0.6$ K for oscillation frequency $f$ determined from a fast Fourier transform (FFT) of the oscillatory torque $\Delta\tau$ is shown in Fig. 2(a). Samples A and B (circles) were measured with pulsed fields up to 65 T while C and D (triangles) were measured in DC fields up to 35 T. The response with $H$ rotated from [001] to [210] ($\theta_1$ rotation) and from [001] to [010] ($\theta_2$ rotation) are similar as shown in the supplementary materials (hereafter we refer to both as $\theta$); we identify 6 branches in the oscillatory pattern with the qualitative behaviors we label as $\alpha_i$, $\beta_i$, $\gamma_i$ ($i = 1,2$). The $\alpha$ group grows rapidly with $\theta$ towards a divergence as $H$ approaches the basal plane, implying that they have a quasi-2D nature (*33*). The value of $f(\theta)$ approaching the $c$-axis $f_0$ for $\alpha_1(\alpha_2)$ is approximately 200 T (930 T), corresponding to a Fermi wave vector $k_F \approx 0.08$ Å$^{-1}$ (0.17 Å$^{-1}$) similar to the inner (outer) Dirac cone areas observed in angle resolved photoemission spectroscopy (ARPES) at K and K' (*8*). We therefore identify these $\alpha$ bands as the quasi-2D massive Dirac bands derived from the Fe kagome network with wave functions primarily confined to the plane. In contrast, the $\beta$ and $\gamma$ groups are free from divergences and instead follow angular dependencies suggestive of three dimensional, closed Fermi sheets, which we identify with the $k_z$-dispersive bands previously reported (*8*).

We examine the $\alpha$ bands in more detail in Fig. 2(b). For an ideal 2D Fermi surface, an evolution $f(\theta) = f_0/\cos\theta$ is expected. This dependence is shown as a dashed line in Fig. 2(b)- we find that the evolution of both of $\alpha$ bands increase more rapidly than this dependence. A deviation of this type is observed in systems with local hyperboloid geometries as depicted in Fig. 2(c) (*10, 33*). However, this is at odds with the lack of $k_z$ dispersion in ARPES (*8*). Moreover, a sinusoidal $k_z$ warping accommodating such a geometry is unable to capture the observed $f(\theta)$ and moreover predicts counterpart extremal frequencies and Yamaji angles (41º and 69º) which are absent (*34*) (see supplementary materials). Interestingly, similar apparently contradictory dHvA spectra were previously observed in elemental ferromagnets (*10, 11, 13*), where it was eventually realized for Ni (*12*) that this could be resolved by considering that spin-orbit coupling would introduce a shift of the energy of the elliptical Fermi pockets up to 50 meV depending on the magnetization direction. Applying such a scenario to the present case of a quasi-2D surface is shown in Fig. 2(d), where *H* plays a dual role setting the direction of magnetization and the plane for cyclotron motion, introducing a faster than $1/\cos\theta$ development. As we describe below, this is well-described by a massive Dirac model with systematically evolving band parameters constructed for the inner Dirac band and extended to capture the outer Dirac band (solid lines in Fig. 2(b)).

We note that from these observations that the size of the Fermi surface can be used as a caliper to probe the orientation of $\vec{m}$. Focusing on the smaller Dirac surface, from ARPES performed between 90 to 110 eV a $k_z$-independent Fermi wave vector is observed corresponding to a circular Fermi surface area $A_k = (0.0259 \pm 0.0008)$ Å$^{-2}$. Converting this to an angular projected area, we find it corresponds to the *c*-normal Fermi surface in Fig. 2(b) at $\theta = (43 \pm 2)$º or a *z*-component of the magnetic moment $m_z \approx 0.7|\vec{m}|$. Studies of bulk magnetic order in Fe$_3$Sn$_2$ have reported a spin-reorienation of the moments towards the basal plane with varying degrees of *c*-axis moment at *T* = 20 K (at which ARPES was performed) accompanied by a variety of magnetic orders including collinear (*35*), non-collinear (*29*), and spin-glass (*36*), while surface probes have suggested that such a reorientation may be first order and partial or complete depending on cooling history (*36*). The acute dependence of $f(\theta)$ to the ferromagnetic order observed here offers a unique window to map the orientation of the $\vec{m}$ for comparison to other surface or bulk sensitive experiments.

To further examine the orientational effect of $\vec{m}$ on the electronic structure, we have measured the effective mass $m^*$ of the Dirac bands as a function of $\theta$. A typical measurement of the dHvA oscillation amplitude ($\theta = 37°$) as a function of $T$ for both $\alpha$ bands is shown in Fig. 3(a) with the double Dirac structure shown as inset. The overall dHvA oscillation amplitude of multiple oscillations in magnetization can be written as (*33*)

$$M_{osc} = \sum_i A_i B^{1/2} R_T^i R_D^i R_S^i \cos\left[2\pi\left(\frac{f_i}{B} + \gamma_i\right)\right] \quad (1)$$

where $i$ is the band index, $A_i$, $f_i$ and $\gamma_i$ are the initial amplitudes, oscillation frequencies and phase factor of the $i$-th band, respectively. $R_T^i = \frac{2\pi^2 k_B T m_i^*}{\hbar eB} \sinh^{-1}\left(\frac{2\pi^2 k_B T m_i^*}{\hbar eB}\right)$ represents the thermal damping factor, $R_D^i = \exp\left[-\frac{2\pi^2 k_B T_D^i m_i^*}{\hbar eB}\right]$ is the Dingle damping factor induced by residual impurities where $T_D$ is the Dingle temperature, $k_B$ the Boltzmann constant, and $2\pi\hbar$ the Planck constant. $R_S^i$ is the modulation due to interfering up and down spin oscillations with spin splitting induced by the magnetic field taken here to be unity given the ferromagnetic spin splitting in excess of 1 eV (*8*). Fitting $R_T^i$ and using the mean inverse field of the FFT window $\bar{B}^{-1} = \frac{1}{2}(B_{min}^{-1} + B_{max}^{-1})$ ($\bar{B} = 30$ T), we find the $\alpha_1$ oscillation has effective mass of $(0.59 \pm 0.04)\, m_e$ while the $\alpha_2$ pocket has a mass of $(2.5 \pm 0.3)\, m_e$.

Extending this analysis across the dHvA spectrum (see supplementary materials), we plot the observed effective masses versus $f$ in Fig. 3(b). We see a monotonic increase of $m^*$ with $f$, but interestingly the ratio $m^*/f$ for $\alpha_1$ is weakly dependent on $\theta$ (see Fig. 3(b) inset). As rigid ellipsoidal (*33*), hyperboloid (*10*), or quasi-2D pockets (*37*) would have a constant ratio, this further suggests the use of a model with a Fermi surface that itself evolves with $\theta$. Based on previous observations of the double massive Dirac spectrum in this system (see schematic in Fig. 3(a)), we analyze the dHvA spectrum with a massive Dirac model. We note that the outer Dirac pocket has been observed to have substantial warping near the Fermi level $E_F$ (illustrated in the inset in Fig. 3(a) and observed with the rapidly growing $m^*(\theta)$ shown in Fig. 3(c) inset); we focus the model on the inner Dirac pocket and approximate the outer Dirac pocket as a copy of this band shifted by the observed $\Delta E = 110$ meV (*8*). In analogy to the spin-orbit models of Ni (*12*), we consider that the Dirac band parameters are modulated by $\vec{m}$. We take the Fermi level (defined from the Dirac point) to be $E_F = \sqrt{(\hbar v_D k_F)^2 + (\Delta/2)^2}$, where $v_D$ is the Dirac velocity, $k_F$ the Fermi wave-vector, and $\Delta$ the Dirac gap. We can then express $f$ and $m^*$ (shown in Fig. 3(c)) as

$$f = \frac{E_F^2 - (\Delta/2)^2}{2e\hbar v_D^2 \cos\theta}, \qquad m^* = \frac{E_F}{v_D^2 \cos\theta} \qquad (2)$$

where $\Delta$, $E_F$ and $v_D$ are $\theta$ dependent band parameters ($\cos\theta$ is the geometric factor associated with the tilted magnetic field). In the $\Delta \to 0$ limit, such models have been previously applied to graphene to successfully describe the disappearing cyclotron mass at charge neutrality (*38, 39*). Assuming a Kane-Mele spin-orbit coupling with massive Dirac fermions (*30*), the intrinsic anomalous Hall conductivity per kagome bilayer provides a further constraint to these parameters

$$\sigma_{xy}^A t = \frac{e^2}{2h}\left(\frac{\Delta}{E_F} + \frac{\Delta}{E_F + \Delta E}\right) \qquad (3)$$

Here $t$ is the thickness of a structural unit that contains a single kagome bilayer. The room temperature $\sigma_{xy}^A$ (Fig. 3(d)) is dominated by the Berry-curvature induced response (*8*); we use this along with $f$ and $m^*$ to quantify the three independent band parameters within this simplified model.

We show the directly calculated $\Delta$, $E_F$, and $v_F$ in Fig. 3(e) along with a schematic band model inset. Generally, all three parameters are suppressed with increasing $\theta$ which can be reasonably captured by polynomials in cosine of the form $A(\theta) = A_0 + A_1\cos\theta + A_2\cos^2\theta$ ($A_i > 0$). With these smooth functions we obtain the solid fits to Fig. 3(c) and (d). In the $\theta \to 0$ limit, we estimate $\Delta_0 = 32$ meV, $v_D^0 = 2.2 \times 10^5$ m/s and $E_F^0 = 112$ meV for the dispersions with moment along the *c*-axis. An extrapolation of the model suggests that for the moment near the basal plane a total shift in $E_F \sim 50$ meV of comparable scale to that reported in Ni (*12*). $\Delta$ shows a stronger reduction while $v_D$ decreases by 34% up to 50°, suggesting increased correlation of the Dirac states for moments in the plane. We can use the band parameters to reconstruct the trends observed in experiment for the inner Dirac surface (solid curves in Fig. 2(b), Fig. 3(b,c,d)); while the outer Dirac surface is beyond our model, we find that a simple scaling of *f* from the inner Dirac surface by a factor of 4.9 approximately captures its angular evolution (see Fig. 2(b)). We note that for $\theta = 42°$ we infer $\Delta_0 = 18$ meV, $v_D^0 = 1.67 \times 10^5$ m/s and $E_F^0 = 95$ meV, in reasonable agreement with the band parameters observed in ARPES (*8*), particularly considering the simplicity of the present model. These observations suggest that in the presence of spin-obit coupling the Dirac bands have a considerable response to changes in the intrinsic ferromagnetism (*9*). Extending the angular range of these measurements as well as more sophisticated modeling of this behavior including the role of the other electronic bands as charge reservoirs are of considerable interest.

Finally, we return to the overall magnetic torque behavior with $H$. In Fig. 4(a) we show the low field torque response for different $\theta$ measured up to 9 T in a superconducting magnet at $T = 3$ K. Similar to the response to high field pulses, for small $\theta$ a sharp kink appears followed by a gradual decay, which evolves to a broader shoulder at larger $\theta$. Despite the apparent qualitative distinction in the torque profiles at small and large $\theta$, all the corresponding $M_T$ curves (Fig. 4(a) inset) behave similarly, showing an initial sharp growth consistent with the soft ferromagnetic nature (*29*) followed by a long tail as a function of field in various angles following primarily $(\mu_0 H)^{-1}$. The latter corresponds to constant $\tau$ expected when when $\vec{m}$ is effectively saturated along $H$ (*40*). This trend is clearer when extended to high field: Fig. 4(b) shows $M_T$ up to 60 T, showing that it is a good approximation that the moment direction is fixed to the applied field (with deviation less than < 0.1° or 0.01 $\mu_B$/f.u.) above 20 T, thus decoupling the evolution of $\vec{m}$ along with the band structure at low fields from the high field regime in which the dHvA oscillations are observed. Quantitatively, from the angular dependence of the torque a moderate easy-plane anisotropy can be inferred (see supplementary materials), similar to previous reports in which shape anisotropy plays an important role (*41*). Further study of the interplay of bulk, surface, and shape anisotropies with the electronic structure of this system is an important area for future work; as the Dirac mass itself can influence magnetic order in similar systems (*42, 43*), an exciting prospect is that the Dirac fermions themselves along with spin-orbit coupling play a role in determining evolution of the magnetic order.

The dHvA results presented here are a thermodynamic probe of the ground state of the correlated, topological bands of $Fe_3Sn_2$. The magnetoquantum oscillations confirm the bulk nature of the quasi-2D massive Dirac bands arising from the kagome network previously observed spectroscopically (*8*) and provide guidance for theoretical models of this system. Viewed more broadly, the results here demonstrate how topological electronic bands can be wed with the robust ferromagnetism in correlated electron systems. Given the widespread use of 3$d$ ferromagnets in spintronics, this provides the exciting prospect that topologically non-trivial analogs of the workhorse materials for spintronics may be developed, allowing direct integration of topological electronic states in to such architectures (*44*). The development of such materials where the charge, spin and heat transport properties are dominated by the topological bands and controllable with spintronic techniques will be an important direction in realizing the promise of topological electronic states to impact the next generation of electronic devices.

## Methods

**Crystal growth and characterization**

Single crystals were grown with an $I_2$ catalyzed reaction starting from Fe and Sn powders. Details of the crystal growth are described in *(8)*.

**High Magnetic Field Measurements**

Piezo torque magnetometry measurements were performed in the National High Magnetic Field Laboratory (NHMFL) at both the DC field (Tallahassee, Florida) and pulsed field (Los Alamos National Laboratory, LANL) facilities. Measurements in the DC field up to 35 T was performed with PRC-400 (Seiko) cantilevers *(45)* in $^3$He atmosphere using the standard lock-in technique with 50 mV AC excitation voltage (~10-20 Hz) to the bridge circuit. Measurements in the pulsed field up to 65 T were performed using PRC-120 (Seiko) cantilevers *(45)* at LANL in both $^3$He and $^4$He atmosphere with a typical high frequency (~ 300 kHz) AC excitation current ~297 $\mu$A. We have repeated the measurements and compared the oscillation amplitudes in $^4$He gas at 4 K with different currents to confirm that this measurement current does not induce significant heating. Temperatures between 1.5 – 4 K were taken with sample immersed in $^4$He liquid.

In both experiments we used a balanced Wheatstone bridge between the piezoresistive pathways with and without the sample to eliminate contributions from the temperature and magnetic field dependence of the piezoresistor to the torque signal *(46)*. Crystals were mounted with $c$ axis perpendicular to the cantilever plane and piezo cantilever arm perpendicular ($\theta_1$ rotation) or parallel ($\theta_2$ rotation) to the hexagonal edge. We converted the measured voltage signal to magnetic torque using the following conversion $\tau = \Delta V/(5.2 \times 10^6 V_0)$ N · m suggested in Ref. *(46)*. Here $\Delta V$ refers to the voltage difference between the two bridge points, and $V_0$ stands for the excitation voltage to the bridge circuit.

**Capacitive Torque measurements**

Low field torque measurements were performed in a commercial superconducting magnet using 10-25 μm Cu:Be foil and signal was acquired with Andeen-Hagerling 2500 AC capacitance bridge. The crystal was attached to the cantilever foil with H20E silver epoxy to prevent detachment in the magnetic field. A typical value of the zero field capacitance is 0.68 pF at $T$ = 3 K in the $^4$He atmosphere.

**Electrical Transport measurements**

The angular-dependent anomalous Hall effect was measured with the standard five probe method using a typical AC excitation current of 2 mA. Both current and voltage leads are placed within the kagome basal plane with current along the [010] direction. The sample was rotated in the magnetic field with $H$ approaching from the $c$-axis ([001]) to the [210] direction (the angular behaviors observed when $H$ is rotated from c-axis to the [010] current direction are similar). The anomalous Hall effect was estimated for each angle as the zero field extrapolation from the linear high field response.

**Acknowledgments**

We are grateful to A. Shekhter for discussions. This research was funded in part by the Gordon and Betty Moore Foundation EPiQS Initiative, grant GBMF3848 to J.G.C. and NSF grant DMR-1554891. L.Y. acknowledges support by the Tsinghua Education Foundation. M.K.C. and R.D.M. were supported by DOE-BES Science at 100T program. M.K. acknowledges a Samsung Scholarship from the Samsung Foundation of Culture. J.L. acknowledges financial support from the Hong Kong Research Grants Council (Project No. ECS26302118). L.Y., M.K., R.C., L.F. and J.G.C. acknowledge support by the STC Center for Integrated Quantum Materials, NSF grant number DMR-1231319. A portion of this work was performed at the National High Magnetic Field Laboratory, which is supported by the National Science Foundation Cooperative Agreement No. DMR-1157490 and DMR-1644779, the State of Florida and the U.S. Department of Energy.

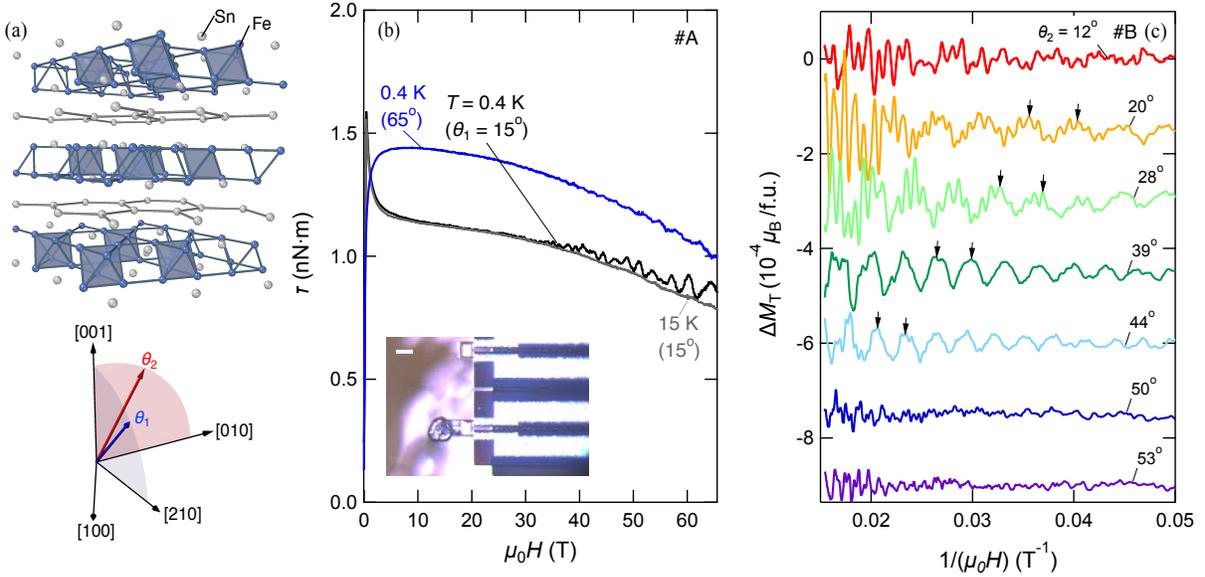

**Figure 1 Pulsed Field Torque Magnetometry and de Haas-van Alphen oscillations in $Fe_3Sn_2$**
(a) Three dimensional crystal structure of $Fe_3Sn_2$ showing the Fe kagome bilayers partitioned by stanene honeycomb layers. The blue clusters are defined by the shortest Fe-Fe bonds (< 2.55 Å). The lower panel illustrates the rotation of magnetic field from out-of-plane to two inequivalent in-plane principal directions. The angles between the field and $c$ axis are defined as $\theta_1$ and $\theta_2$ in the two rotation planes, respectively. (b) Magnetic torque $\tau$ measured up to 65 T for $\theta_1 = 15°$ and $65°$ with de Haas-van Alphen oscillations observed above ~ 20 T for $T = 0.4$ K. The inset shows an optical image of the piezoresistive cantilever with one crystal of hexagonal, plate-like $Fe_3Sn_2$ (the scale bar is 50 μm). (c) Oscillatory part of the transverse magnetization $\Delta M_T$ at selected angles at base temperature $T = 0.5$-$0.6$ K versus inverse magnetic field. The black arrows correspond to the 8th and 9th oscillation of the slow frequency at each angle.

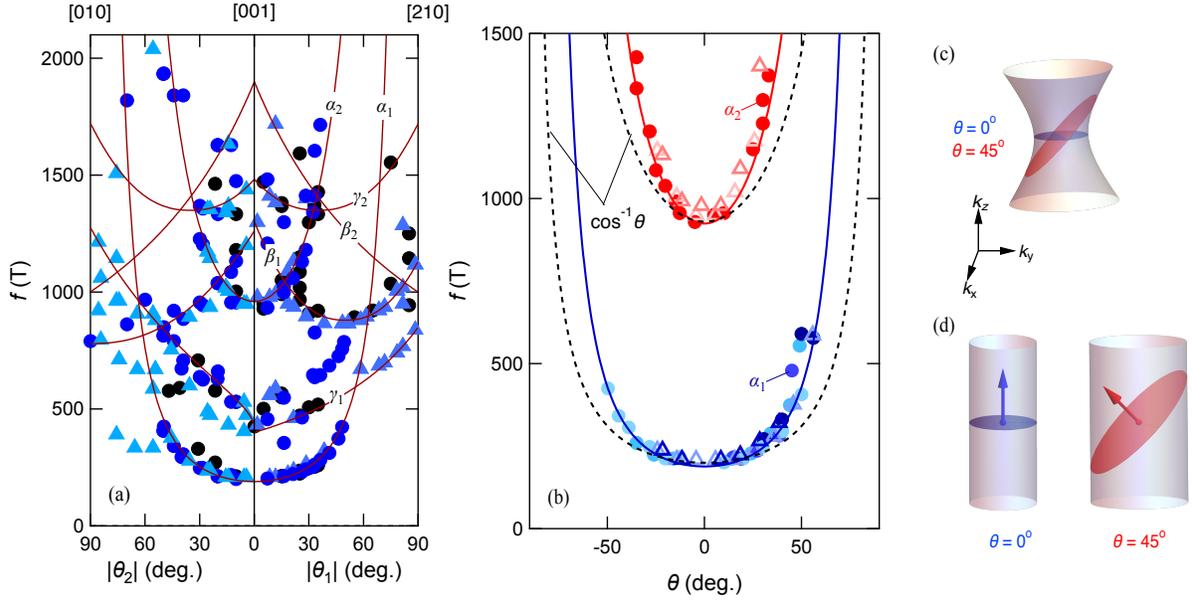

**Figure 2 Angular dependence of dHvA oscillation frequencies** (a) Angular dependence of all Fast Fourier Transform (FFT) frequencies with rotation from [001] to [010] (left panel) and from [001] to [210] direction (right panel). Solid circles are collected from pulsed field experiments while solid triangles are from DC field experiments. Data taken from different samples are represented with different colors. The dark red curves are guides to the eye. (b) Angular dependence of $\alpha_1$ and $\alpha_2$ pockets. The dashed lines are the behavior expected for a 2D cylindrical Fermi surface ($1/\cos\theta$) and the solid lines are a massive Dirac model (see text). (c) Schematic of a hyperboloid Fermi surface whose smallest extremal area evolves faster than $1/\cos\theta$ with rotating magnetic field. (d) Schematic of quasi-2D Fermi surface where the $k_z$-dispersionless Fermi wavevector changes with the direction of magnetization (shown as arrows).

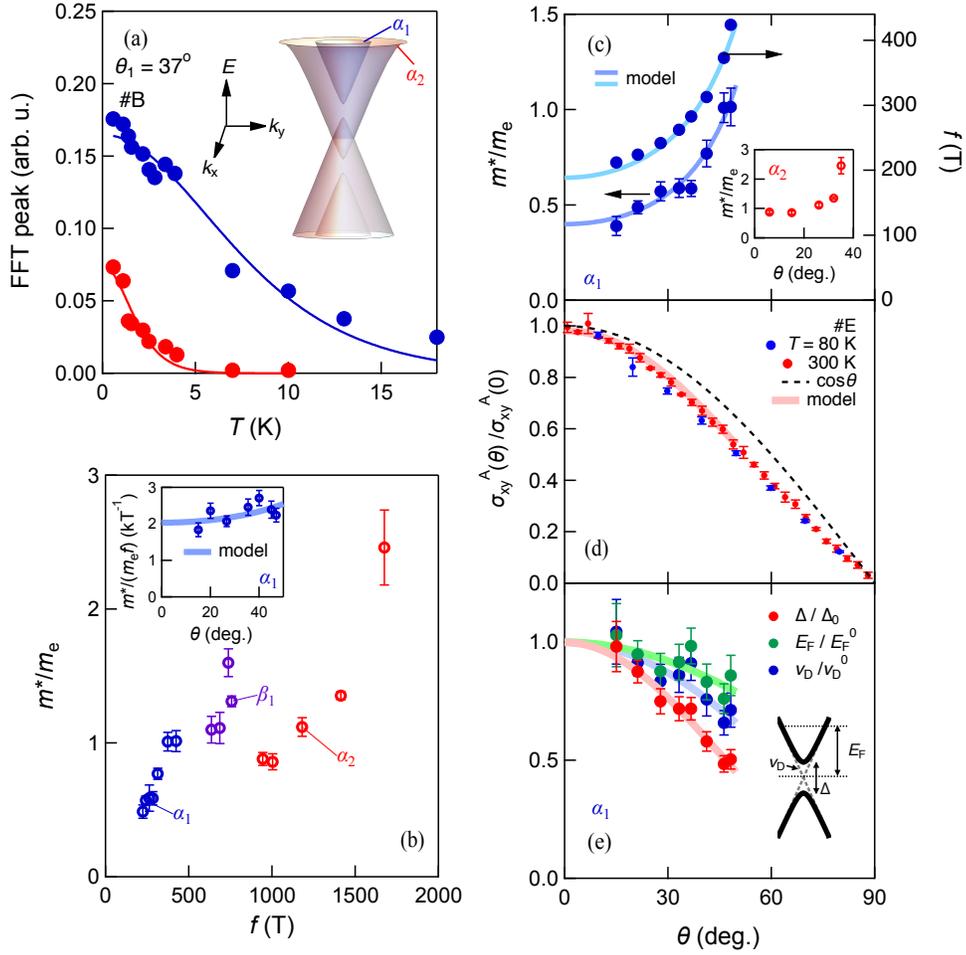

**Figure 3 Massive Dirac model of de Haas-van Alphen effect in Fe$_3$Sn$_2$** (a) Temperature dependence of oscillation amplitude and Lifshitz-Kosevich fitting of $f$ = 1675 T ($\alpha_2$) and 283 T ($\alpha_1$) at $\theta_1 = 37°$. Inset shows a schematic of the double Dirac spectrum. (b) The observed effective mass $m^*/m_e$ versus oscillation frequency $f$ for observed Fermi pockets. The inset shows the angular dependence of the ratio $m^*/m_e f$ for $\alpha_1$ along with the massive Dirac model (see text). (c) Angular dependence of $m^*/m_e$ and $f$ for the inner Dirac pocket (outer $m^*/m_e$ pocket shown inset), and (d) anomalous Hall conductivity $\sigma_{xy}^A$ normalized to the zero angle value ($\sigma_{xy}^A(0)$ = 130 $\Omega^{-1}$cm$^{-1}$ at 300 K and $\sigma_{xy}^A(0) = 169$ $\Omega^{-1}$cm$^{-1}$ at 80 K), respectively, with solid curves showing the massive Dirac model (see text). (e) Angular dependence of the massive Dirac band parameters where the gap is normalized to $\Delta_0$ = 32 meV, the Dirac velocity normalized to $v_D^0$ = $2.2 \times 10^5$ m/s and the Fermi energy is normalized to $E_F^0$ = 112 meV with schematic Dirac band shown in inset.

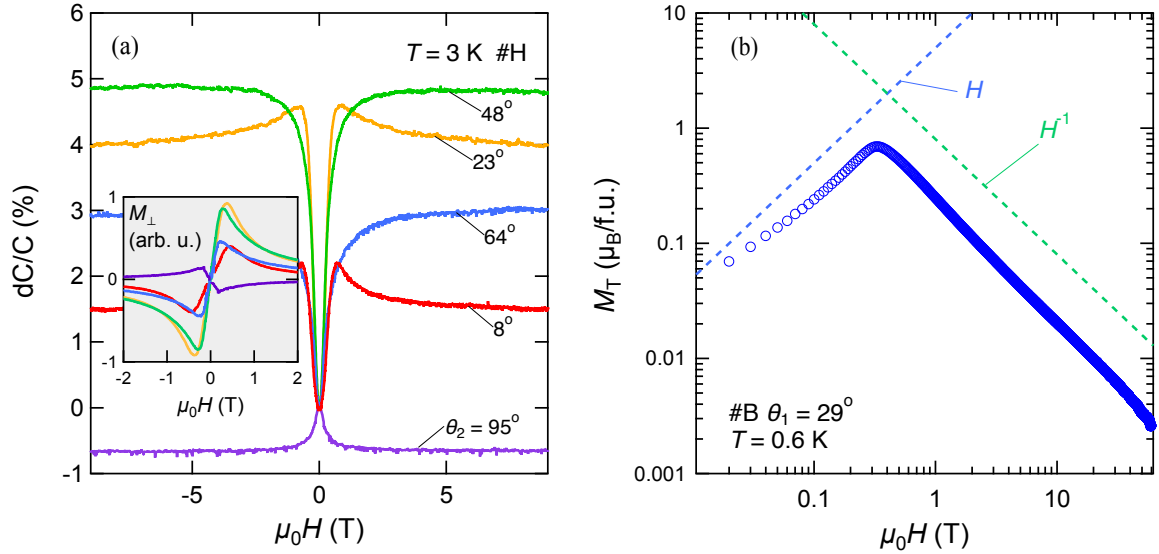

**Figure 4 Torque response from the soft ferromagnetism in Fe$_3$Sn$_2$** (a) Low field magnetic torque at selected angles at $T = 3$ K measured with capacitive cantilever in a superconducting magnet. At low angles the torque response exhibits an initial increase which gradually transforms to a broad shoulder at high angles, consistent with the observation at high fields with piezoresistive cantilevers. The inset shows the transverse magnetization extracted for each torque curve. (b) Pulsed field transverse magnetization $M_T$ up to 60 T at $\theta_1 = 29°$ at $T = 0.61$ K shown in a log-log scale. $M_T$ attains a maximum approximately 0.7 $\mu_B$ per formula unit below 1 T and at higher fields follows an approximately $H^{-1}$ dependence.